\documentclass[12pt,paper,onecolumn]{IEEEtran}

\usepackage{amsmath}
\usepackage{amssymb}
\usepackage{latexsym}
\usepackage{amsfonts}

\begin{document}
\pagestyle{headings}  
\vspace{1cm}

%
\title{On the $k$-error Linear Complexity  for $p^n$-periodic Binary Sequences  via Hypercube Theory}

\author{
\authorblockN{Jianqin Zhou\\}
\authorblockA{Department of Computing, Curtin University, Perth, WA 6102 Australia\\
Computer Science School, Anhui Univ. of
Technology, Ma'anshan, 243002 China\\
 \  zhou9@yahoo.com\\
%
\ \\
Wanquan Liu\\
Department of Computing, Curtin University, Perth, WA 6102 Australia\\
W.Liu@curtin.edu.au\\
\ \\
Guanglu Zhou\\
Dept of Mathematics \& Statistics, Curtin University, Perth, WA 6102 Australia\\
G.Zhou@curtin.edu.au
 }
}
\maketitle              

\begin{abstract}
The linear complexity and the $k$-error linear complexity of a
binary sequence are important security measures for key stream
strength. By studying  binary sequences with the minimum Hamming
weight, a new tool named as hypercube theory is developed for
$p^n$-periodic binary sequences. In fact, hypercube theory is based
on a typical sequence decomposition and it is a very important tool
in investigating the critical error linear complexity spectrum
proposed by Etzion et al. To demonstrate the importance of hypercube
theory, we first give a standard hypercube decomposition based on a
well-known algorithm for computing linear complexity and show that
the linear complexity of the first hypercube in the decomposition is
equal to the linear complexity of the original sequence. Second,
based on such decomposition, we give a complete characterization for
 the first decrease of the linear complexity for a $p^n$-periodic binary sequence $s$. This
significantly improves the current existing results in literature.
As to the importance of the hypercube, we finally derive a counting
formula for the $m$-hypercubes with the same linear complexity.


\end{abstract}

\section{Introduction}


The linear complexity of a sequence  $s$, denoted as
$L(s)$, is defined as the length of the
shortest linear feedback shift register (LFSR) that can generate the
sequence. The concept of linear complexity is very useful in the
study of security of stream ciphers for cryptographic applications
\cite{Ding,Games}. In fact,  a high linear complexity is  necessary  for the security
of a key stream. However, high linear complexity can not
 guarantee a sequence is definitely secure. For example, if a small
number of changes to a sequence can greatly reduce its linear
complexity, then the resulting key stream would be cryptographically
weak. To tackle this issue, Ding, Xiao and Shan \cite{Ding} proposed
the weight complexity and sphere complexity. Stamp and Martin
\cite{Stamp} introduced the  $k$-error linear complexity, which is
very similar to the sphere complexity. Specifically, suppose that
$s$ is a sequence  with period $N$, for any $k(0\le k\le N)$, the
$k$-error linear complexity of $s$, denoted as $L_k(s)$,  is defined
as the smallest linear complexity when any $k$ or fewer terms of the
sequence are changed within one period.

One important result, proved by Kurosawa et al.  \cite{Kurosawa}, is
that the minimum number $k$ for which the $k$-error linear
complexity of a $2^n$-periodic binary sequence $s$ is strictly less
than the linear complexity $L(s)$ of $s$ is determined by
$k_{\min}=2^{W(2^n-L(s))}$, where $W(a)$ denotes the Hamming weight
of the binary representation of an integer $a$. For a $p^n$-periodic
binary sequence, where $p$ is an odd prime and 2 is a primitive root
modulo $p^2$, Meidl \cite{Meidl} studied the minimum value $k$ for
which the $k$-error linear complexity is strictly less than the
linear complexity. Han et al. \cite{Han} investigated the same issue
in a new viewpoint different from the approach by Meidl
\cite{Meidl}. Currently, the best result on this smallest $k$
\cite{Meidl} is characterized with an upper bound.  In this paper,
we derive a precise formula for such smallest $k$ using the proposed
hypercube theory. This is one main contribution of this paper.

On the other hand, Etzion et al. \cite{Etzion} studied the error
linear complexity spectrum of binary sequences with period $2^n$.
Etzion et al. gave a precise categorization of those sequences with
the $k$-error linear complexity equal to linear complexity or zero,
as well as an enumeration of these sequences. In fact, for the error
linear complexity spectrum of binary sequences with period $p^n$, it
is very hard to study the second decrease point for the linear
complexity and we obtain a fundamental result on this issue in this
paper by using the standard hypercube decomposition.

As a small number of element changes in a sequence may lead to a
sharp decline of its linear complexity. Therefore we really need to
study  stable sequences in which even a small number of element
changes do not reduce their linear complexity. The stable $k$-error
linear complexity is introduced  hence to deal with this problem as
follows. Suppose that $s$ is a sequence over $GF(2)$ with period
$N$. For $k(0\le k\le N)$, the $k$-error linear complexity of $s$ is
defined as stable when any $k$ or fewer  terms of the sequence are
changed within one period, the linear complexity does not decline.
In this case, the $k$-error linear complexity of sequence $s$ is
equivalent to its linear complexity. By using Theorem 4.1 in this
paper, we find a way to construct such stable sequence over $GF(2)$
with period $N$. Also the second critical point for a hypercube is
also fully characterized in this paper.

As to the importance of the hypercube defined in this paper, we
derive a counting formula for m-hypercubes with the same linear
complexity and this will pave a way for other applications of
hypercube in future.

Technically, the results related to $k$-error linear complexity  for
$2^n$-periodic binary sequences
\cite{Etzion,Kurosawa,Lauder,Meidl2005,Stamp,Zhou_Liu,Zhu} are
mainly based on
 the Games-Chan algorithm
\cite{Games} which efficiently computes the linear complexity of
$2^n$-periodic binary sequences. In contrast, those for
$p^n$-periodic
 sequences \cite{Han,Meidl}  are mainly based on the XWLI
algorithm given by Xiao, Wei, Lam, and Imamura \cite{Xiao}, which
efficiently computes the linear complexity of $p^n$-periodic
sequences. Generally, the latter is more complex to study. For easy
understanding, we put the corresponding results for the period $2^n$
binary sequence briefly in this paper.

The Cube Theory is  introduced   in \cite{Zhou_Liu2013} to study the
$k$-error linear complexity of  $2^n$-periodic binary sequences.
Similarly, by studying  sequences with the minimum Hamming weight, a
new tool called hypercube theory is developed in this paper for
$p^n$-periodic binary sequences.
 We first  give a
general hypercube decomposition approach. Second,  a characterization is presented about
 the first decrease  in the
$k$-error linear complexity for a $p^n$-periodic binary sequence $s$ based on hypercube theory.
This significantly improves one theorem  in \cite{Meidl}.
 One significant benefit for the standard hypercube decomposition is for us to construct sequences with
the maximum stable $k$-error linear complexity.
 Also the second decrease point for linear complexity of a $p^n$-periodic binary sequence is also investigated with some novel results. Finally, a counting
formula for $m$-hypercubes with the same linear complexity is
derived.

%

The rest of this paper is organized as follows. In  Section II, some
preliminary results are presented. In  Section III, we will
introduce the hypercube decomposition  and investigate the linear
complexity. Some main results on $k$-error linear complexity are
presented in Section IV. The conclusions are given in Section V.

\section{Preliminaries}

For definitions and notations not presented here, we follow \cite{Meidl}.
In this section we give some preliminary results which will be used in the sequel.

Let $x=(x_1,x_2,\cdots,x_n)$ and
$y=(y_1,y_2,\cdots,y_n)$ be vectors over $GF(q)$, define
$$x+y=(x_1+y_1,x_2+y_2,\cdots,x_n+y_n).$$
If $q=2$,  $x+y$ is identical to  $x\bigoplus y$.

The Hamming weight of an $N$-periodic sequence $s$ is defined as the
number of nonzero elements  per period of $s$, denoted by
$W_H(s)$.  The distance of two elements is defined as the
difference of their indexes. Specifically, for  an $N$-periodic sequence $s=\{s_0, s_1, s_2,
\cdots, s_{N-1}\}$, the distance of $s_i,s_j$ is $j-i$, where $0\le i\le j\le N$.

Let $q$ be a primitive root modulo $p^2$ and $s$ a $p^n$-periodic  sequence over $GF(q)$. Denote
$$s^{(n)}=\{s_0^{(n)}, s_1^{(n)}, s_2^{(n)},
\cdots, s_{p^n-1}^{(n)}\}$$  as a period of $s$. The linear
complexity $L(s)$ of a $p^n$-periodic sequence $s$ can be
efficiently obtained by the following XWLI algorithm
\cite{Han,Xiao}.

\noindent {\bf Algorithm 2.1} XWLI Algorithm: Initially set $l=0, L=0$.
Let $a=\{s_{0}, s_{1},
\cdots, s_{p^{n}-1}\}$. We  divide $a$ into $p$  parts with
$A_i=\{s_{ip^{n-1}}, s_{ip^{n-1}+1},
\cdots, s_{(i+1)p^{n-1}-1}\}$, $0\le i<p$, and $a=\{A_0,A_1,\cdots, A_{p-1}\}$.

For $l<n$

(i) If $A_0=A_1=\cdots= A_{p-1}$ then $a\leftarrow A_0$ and $l\leftarrow l+1$.

(ii) Otherwise, $a\leftarrow A_0+A_1+\cdots+ A_{p-1}, l\leftarrow l+1, L\leftarrow L+(p-1)p^{n-l}$.

For $l=n$

if $a\neq \{0\}$, then $L\leftarrow L+1$, end if.

Stop.

Finally, we have that $L(s) = L$.

\

 We have the following observations for {\bf Algorithm 2.1}.

 First, in the $l$th step, the length of each $A_i$ is $p^{n-l}$.
  Further if the position difference
 of two non-zero elements of sequence $a$ is $(px+i)p^{n-l}$,
where  $i, x$ and $y$ are non-negative integers, and $0<i<p$. Then
the two non-zero elements  must be  in two different  $A_{i_1}$ and
$A_{i_2}$, so these two nonzero elements can be removed or reduce to
one non-zero element after the $l$th step operation.

Second, the
operation in (i) will not change its linear complexity.

Third, in
the end of {\bf Algorithm  2.1}, if $a= \{0\}$, then there must
exist $l_1$, such that in the $l_1$th step, $A_0+A_1+\cdots+
A_{p-1}=\{0,0,\cdots ,0\}$, but
 $\{A_0,A_1,\cdots, A_{p-1}\}\ne \{0,0,\cdots ,0\}$.

{\bf Remark 2.1} Assume that in the $k$th step, $1\le k\le n$,
$A_0=A_1=\cdots= A_{p-1}$ are not true, the
 linear complexity is increased by $(p-1)p^{n-k}$. Then after the $k$th step,
  the sum of all possible  linear complexity increase is  $$(p-1)p^{n-k-1}+(p-1)p^{n-k-2}+\cdots+(p-1)p +(p-1)+1=p^{n-k}<(p-1)p^{n-k}$$
This implies that in {\bf Algorithm 2.1}, the first increase of
linear complexity is bigger than the sum of all possible latter
increase. This is an important property for the {\bf Algorithm 2.1}.

 The above observations will help us to understand the hypercube
 greatly.

\

Let $q$ be a primitive root modulo $p^2$ and $s$ a $p^n$-periodic  sequence over $GF(q)$. Han, Chung  and Yang \cite{Han} showed that
 the linear complexity of $s$
can be expressed as
\begin{eqnarray}
L(s)=\epsilon +(p-1)\sum\limits_{v\in V} p^{v-1}
\end{eqnarray}
where $V\subseteq\{1,2,\cdots, n\}$
and $\epsilon\in\{0,1\}$.


The following lemma is a well known result on the number of the
$p^n$-periodic binary sequences with a given linear complexity.

\noindent {\bf Lemma  2.1} (\cite{Meidl2002}): Let $q$ be a primitive root modulo $p^2$ and $s$ a $p^n$-periodic  sequence over $GF(q)$
with linear complexity $L(s)=\epsilon +(p-1)\sum\limits_{v\in V} p^{v-1}$.
Then the number of $p^n$-periodic sequence $s$ with linear complexity $L(s)$ is given by
\begin{eqnarray}N(L(s))=\prod\limits_{v\in V}(q^{(p-1)p^{v-1}}-1)
\end{eqnarray}
where $V\subseteq\{1,2,\cdots, n\}$
and $\epsilon\in\{0,1\}$.

 Sequence decomposition plays an important role for linear
complexity investigation \cite{Zhou_Liu}. We first present a lemma
for the linear complexity of the sum of two $2^n$-periodic binary sequences.

\noindent {\bf  Lemma 2.2} (\cite{Zhou_Liu}):  Let $s_1$ and $s_2$ be two binary sequences
with period $2^n$. If $L(s_1)\ne L(s_2)$, then
$L(s_1+s_2)=\max\{L(s_1),L(s_2)\} $.

However the above property does not hold any more for $p^n$-periodic
binary sequences as demonstrated by  the following three examples.
These imply that in terms of the linear complexity, $p^n$-periodic
binary sequences is much more complex than $2^n$-periodic sequences.

{\bf Example 2.1} Let $s_1=\{100\ 100\ 100 \}$ and $s_2=\{010\ 000\ 000\}$, where $q=2, p=3$.
Then $L(s_1)=3, L(s_2)=2\times3 +2+1=9$ and $L(s_1+s_2)=2\times3 +2=8$. So, $\min\{L(s_1),L(s_2)\}<L(s_1+s_2)<\max\{L(s_1),L(s_2)\} $.

{\bf Example 2.2} Let $s_1=\{100\ 100\ 100 \}$ and $s_2=\{110\ 000\ 000\}$, where $q=2, p=3$. Then $L(s_1)=3, L(s_2)=2\times3 +2=8$ and $L(s_1+s_2)=2\times3 +2+1=9$. Hence, $L(s_1+s_2)>\max\{L(s_1),L(s_2)\} $.

{\bf Example 2.3} Let $s_1=\{111\ 000\ 000 \}$ and $s_2=\{000\ 111\ 111\}$, where $q=2, p=3$. Then $L(s_1)=2\times3 +1=7, L(s_2)=2\times3 =6$ and $L(s_1+s_2)=1$. Hence, $L(s_1+s_2)<\min\{L(s_1),L(s_2)\} $.

The Cube Theory is  introduced   in \cite{Zhou_Liu2013} to study  the $k$-error linear complexity of  $2^n$-periodic
binary sequences. In a similar approach,  we will present Hypercube Theory to  study  the $k$-error linear complexity of  $p^n$-periodic
binary sequences  in next section.
 Therefore, cube theory and some related results are presented here as preliminaries. For further discussions about  cube theory, please refer to \cite{Zhou_Liu2013}.


Suppose that the position difference
 of two non-zero elements of $2^n$-periodic
binary sequence $s$ is $(2x+1)2^y$, where  $x$ and $y$ are
non-negative integers. From  the Games-Chan algorithm \cite{Games}
which divides the sequence to half in each step,  only in the
$(n-y)$th step, the sequence length is $2^{y+1}$, so the two
non-zero elements must be in the left and right half of the sequence
respectively, thus they can be removed or reduce to one non-zero
element. Therefore we have the following definition.

\noindent {\bf Definition  2.1}(\cite{Zhou_Liu2013}): Suppose that the position difference
 of two non-zero elements of $2^n$-periodic
binary sequence $s$ is $(2x+1)2^y$, both $x$ and $y$ are
non-negative integers. Then the distance between the two elements is
defined as $2^y$.

\noindent {\bf Definition  2.2}(\cite{Zhou_Liu2013}): Suppose that $s$ is a binary sequence
with period $2^n$, and there are $2^m$ non-zero elements in $s$, and
$0\le i_1< i_2<\cdots<i_m<n$. If $m=1$, then there are 2 non-zero
elements in $s$ and the distance between the two elements is
$2^{i_1}$, so it is called as a 1-cube. If $m = 2$, and $s$ has 4
non-zero elements which form a rectangle with the lengths of 4 sides being
$2^{i_1}$ and $2^{i_2}$ respectively, so it is called as a 2-cube.
In general, if $s$ has $2^{m-1}$ pairs of non-zero elements, in which
there are $2^{m-1}$ non-zero elements which form a $(m-1)$-cube, the
other $2^{m-1}$ non-zero elements also form a $(m-1)$-cube, and the
distance between each pair of elements are all  $2^{i_m}$, then the
sequence $s$ is called as an $m$-cube. In this case the linear complexity of $s$ is
 called as the linear complexity of the cube as well.

 Cube is a very special sequence with a unique structure. For an ordinary sequence $s$, with the
  standard cube decomposition \cite{Zhou_Liu2013}, $s$ can be decomposed into a series of cubes.

\noindent {\bf Definition  2.3}(\cite{Zhou_Liu2013}):
 A non-zero element  of $2^n$-periodic
binary sequence $s$ is  called a vertex. Two vertices can  form an
edge. If the distance between the two elements (vertices) is $2^y$,
then the length of the edge is defined as $2^y$.

As demonstrated in \cite{Zhou_Liu2013}, the linear complexity of a
$2^n$-periodic binary sequence with only one cube has the following
nice property.

\noindent {\bf Theorem   2.1}(\cite{Zhou_Liu2013}): Suppose that $s$ is a binary sequence
with period $2^n$, and non-zero elements of $s$ form an $m$-cube with
lengths of  edges  $ 2^{i_1}, 2^{i_2},\cdots ,2^{i_m}$ $(0\le i_1<
i_2<\cdots<i_m<n )$ respectively, then
$L(s)=2^n-(2^{i_1}+2^{i_2}+\cdots+2^{i_m})$.

For example, let $s$ be the binary sequence
$\{\overbrace{11\cdots11}^{2^k}0\cdots0\}$. Its period is $2^n$, and
there are only $2^k$ continuous nonzero elements at the beginning of
the sequence. Then one ca prove that it is a $k$-cube  with lengths
of edges $ 2^{0}, 2^{1},\cdots ,2^{k}$, and $L(s)=2^n-(2^{0}+
2^{1}+\cdots +2^{k})$.

\section{Hypercube Theory and Linear Complexity}

Follow the idea of The Cube Theory, we will introduce the hypercube
theory in this section. First present
 some definitions and preliminary results.

Suppose that the position difference
 of two non-zero elements of $p^n$-periodic binary sequence $s$ is $(px+i)p^y$,
where  $i, x$ and $y$ are non-negative integers, and $0<i<p$.
From Algorithm 2.1, only in the $(n-y)$th step, the length of $A_i$ is $p^y$, so the two non-zero elements  must be  in two different  $A_{i_1}$ and $A_{i_2}$, thus they can be removed or reduce to one non-zero element. Therefore we have the following definition.

\noindent {\bf Definition  3.1} Suppose that the position difference
 of two non-zero elements of $p^n$-periodic binary sequence sequence $s$ is $(px+i)p^y$,
where  $i, x$ and $y$ are non-negative integers, and $0<i<p$. Then
the distance between the two elements is defined as  $p^y$.

Next we define the hypercube based on Algorithm 2.1.

\noindent {\bf Definition  3.2}  Let $2$ be a primitive root modulo $p^2$, $s$ a $p^n$-periodic  binary sequence.
When computing the linear complexity of $s$ by Algorithm 2.1, if there is no  decrease of nonzero element  in $s$ in the operation $a\leftarrow A_0+A_1+\cdots+ A_{p-1}, $(except for the last operation), then $s$ is defined as a hypercube. In this case, the linear complexity of sequence $s$ is defined as the linear complexity of hypercube $s$.

For example, let $n=3, p=3$, $s^{(n)}=\{110\ 000\ 000\ 110\ 000\ 000\ 110\ 000\ 000\}$. In  the first operation $a\leftarrow A_0+A_1+\cdots+ A_{p-1}, $ from $\{110\ 000\ 000\}$ to $\{110\}$, there is no decrease of nonzero element numbers. So  $s^{(n)}$ is a hypercube.

 However, let $n=3, p=3$, $s^{(n)}=\{110\ 100\ 100\ 110\ 100\ 100\ 110\ 100\ 100\}$. In  the first operation $a\leftarrow A_0+A_1+\cdots+ A_{p-1}, $ from $\{110\ 100\ 100\}$ to $\{110\}$, two nonzero element disappear. So  $s^{(n)}$ is not a hypercube. Later we will show that the sequence $s^{(n)}$ can be
decomposed into several  hypercubes.

\

\noindent {\bf Definition  3.3} Let $2$ be a primitive root modulo $p^2$, $s$ a $p^n$-periodic  binary sequence.
When computing the linear complexity of $s$ by Algorithm 2.1.

i) (Vertex) At the end of the algorithm,  if $a=\{1\}$, then the
nonzero element   is  called a vertex; if $a=\{0\}$, then there must
exist $l_1<n $, such that in the $l_1$th step, $A_0+A_1+\cdots+
A_{p-1}=\{0,0,\cdots ,0\}$, but
 $\{A_0,A_1,\cdots, A_{p-1}\}\ne \{0,0,\cdots ,0\}$.
  In this case, the $p$-tuple $\{A_0,A_1,\cdots, A_{p-1}\}=B$   is  called a vertex, where $A_i=\{s_{ip^{j}}, s_{ip^{j}+1},
\cdots, s_{(i+1)p^{j}-1}\}$, $0\le i<p, 0\le j<n$. $j$ is defined as
the length of the vertex (There is no length for vertex when
$a={1}$). The original sequence $s$ may include one or more this
defined tuple $B$ (vertex) in its sequence.

ii) (Edge) Two vertices of the same kind ($a={1}$ or $a={0}$) can
form an edge. If the vertex is a nonzero element, then the length of
the edge is defined as the distance of the two nonzero elements; If
the vertex is a nonzero $p$-tuple,
 then the length of the edge is
defined as the distance of the two first  elements in each $p$-tuple.

With above definition of vertex, for a hypercube $s$, if the number of its vertices
 is $p^m$, then hypercube $s$ is called as $m$-hypercube, and
in this case, the dimension of hypercube $s$ is defined as $m$.

For example, let $n=3, p=3$, $s^{(n)}=\{110\ 000\ 000\ 110\ 000\
000\ 110\ 000\ 000\}$. There are 3 vertices of 1 tuple $\{110\}$,
and the length of this tuple  is 0. So $s^{(n)}$ is a $1$-hypercube.

Let $s^{(n)}=\{111\ 111\ 000\ 111\ 111\ 000\ 111\ 111\ 000\}$. There
are 3 vertices with the same tuple $\{111\ 111\ 000 \}$, which has
length 1. So $s^{(n)}$ is a $1$-hypercube.

Hypercube is a very special sequence with a unique structure.
However, based on  Algorithm 2.1, we can develop a standard
hypercube decomposition algorithm, so that any $p^n$ periodic binary
sequence can be decomposed into a series of hypercubes as
demonstrated in the following algorithm.

\noindent {\bf Algorithm 3.1}

\noindent {\bf Input:} $s^{(n)}_0$ is a  binary sequence  with period
$p^n$.

\noindent {\bf Output:} A  hypercube decomposition of sequence $s^{(n)}_0$.

\noindent Step 1.  Let $s^{(n)}=s^{(n)}_0$.  Divide $s^{(n)}$ into
$p$ equal parts, $A_i=\{s_{ip^{n-1}}, s_{ip^{n-1}+1}, \cdots,
s_{(i+1)p^{n-1}-1}\}$, $0\le i<p$, and let $a=\{A_0,A_1,\cdots,
A_{p-1}\}$. While $\{A_0,A_1,\cdots, A_{p-1}\}$ is not a vertex, run
Step 2 and Step 3.

\noindent Step 2.
 If $A_0=A_1=\cdots= A_{p-1}$
 then we consider $A_0$, and let $a\leftarrow A_0$. If we change  $A_0$ in the following steps,  we will also make the same changes to $A_1, A_2,\cdots, A_{p-1}$ in $s^{(n)}$.

\noindent Step 3. Otherwise, $a\leftarrow A_0+A_1+\cdots+ A_{p-1}, $
 then we
consider $  A_0+A_1+\cdots+ A_{p-1}$.

If the number of nonzero elements in $a$ is less than the sum of the
number of nonzero  elements in each $A_i$ for $0\le i<p$,   then we
can change the nonzero elements in $A_0, A_1,\cdots, A_{p-1}$
accordingly (change $A_i$ to $\tilde{A}_i$) such that i)   $
\tilde{A}_0+\tilde{A}_1+\cdots+ \tilde{A}_{p-1} $ is still equal to
the original $a$ and ii)
  the number of nonzero elements in $a$ is the same
as the sum of the number of nonzero elements in $\tilde{A}_i$ for
$0\le i<p$. iii) $\tilde{A}_0=\tilde{A}_1=\cdots= \tilde{A}_{p-1}$
is still not true ( refer to  Appendix 1) on how to achieve it).
These changes will guarantee that the nonzero elements in operation
$ a \leftarrow \tilde{A}_0+\tilde{A}_1+\cdots+ \tilde{A}_{p-1}$ will
not be reduced and this will make sure these changes will follow the
definition of hypercube.

Further, if we change a nonzero element of  $a$ in the following
steps, we will also change its corresponding nonzero element back in
$ \tilde{A}_0,\tilde{A}_1,\cdots,$ or $\tilde{A}_{p-1}$ in
$s^{(n)}$,
 such that  $ a=\tilde{A}_0+\tilde{A}_1+\cdots+ \tilde{A}_{p-1} $.
 This is possible as the number of nonzero element in $a$ equals to
 the sum of nonzero elements in $
\tilde{A}_0,\tilde{A}_1,\cdots,$ and $\tilde{A}_{p-1}$. This will
make sure the changed sequence in each operation can be traced back
though these back changes may not be unique.


\noindent Step 4. Repeat  above operations, until that $a$ is
reduced to one vertex. In above process, keep all possible changes
in $s^{(n)}$ and it will finally become a hypercube $h_1$ with
linear complexity $L(s^{(n)}_0)$. (refer to Appendix ii)).

\noindent Step 5. With $s_0^{(n)}\bigoplus h_1$, where $s_0^{(n)}$
is the original sequence, run Step 1 to Step 4. We can obtain a
hypercube $h_2$ with linear complexity less than $L(s_0^{(n)})$.
(refer to Appendix iii)).

\noindent Step 6. With these nonzero elements  left in $s_0^{(n)}$, run Step 1 to Step 5 recursively
 we will obtain
a series of hypercubes in the descending order of linear complexity.
Finally, we can obtain the following decomposition.
$$s_0^{(n)}=h_1\bigoplus h_2\bigoplus h_3\bigoplus \cdots$$

%
%

\

For the correctness of {\bf Algorithm 3.1}, please see Appendix 1), 2) and 3).

\

The above process is defined  as {\bf the standard hypercube decomposition} for a sequence $s^{(n)}$.

 For example, let $n=3, p=3$, $s^{(n)}=\{110\ 100\ 100\ 110\ 100\ 100\ 110\ 100\ 100\}$.
Then it can be decomposed into 1-hypercube $\{000\ 100\ 100\ 000\ 100\ 100\ 000\ 100\ 100\}$ and 1-hypercube \\ $\{110\ 000\ 000\ 110\ 000\ 000\ 110\ 000\ 000\}$. They have linear complexity 6 and 8, respectively.

\

%

Based on {\bf Algorithm 2.1} and the standard hypercube
decomposition, we first consider the linear complexity of a sequence
with only one hypercube as corresponding to Theorem 2.1 for the case
of $2^n$ periodic sequences. As shown in the following theorem,
there are three kinds of vertices, so the linear complexity of one
hypercube is much more complex than the linear complexity of one
cube in Theorem 2.1.

\noindent {\bf Theorem   3.1} Suppose that $s$ is a binary sequence
with period $p^n$, and  further $s$ is an $m$-hypercube with lengths
of edges being $ p^{i_1}, p^{i_2},\cdots ,p^{i_m}$ $(0\le i_1<
i_2<\cdots<i_m<n )$ respectively. Then
$$L(s)=\epsilon-1+p^n-(p-1)(p^{i_1}+p^{i_2}+\cdots+p^{i_m})$$
where $\epsilon$ has three cases: i) if the vertex of hypercube is a
nonzero element, then $\epsilon=1$; ii)  if the vertex of hypercube
is a nonzero tuple and the length of the vertex is 0, then
$\epsilon=0$; iii) if the vertex of hypercube is a nonzero tuple and
the length of the vertex is $j>0$, then
$\epsilon=(1-p)(p^0+p^1+\cdots+ p^{j-1})$.

\noindent \begin{proof} We prove the three cases separately as
follows.

i) If the vertex is a nonzero element, then $a={1}$ at the end of
{bf Algorithm 2.1}. From {\bf Algorithm 2.1}, in the $k$th step,
$1\le k\le n$, if and only if one period of the sequence can not be
divided into $p$ equal parts, then the
 linear complexity will be increased by $(p-1)p^{n-k}$.  So, the all possible linear complexity increases are:
  $$(p-1)p^{n-1}, (p-1)p^{n-2}, \cdots,(p-1), 1.$$

Assume that  sequence $s$ is a $m$-hypercube with lengths of  edges
being $ p^{i_1}, p^{i_2},\cdots ,p^{i_m}$ $(0\le i_1<
i_2<\cdots<i_m<n )$ respectively, then we will prove that the linear
complexity of this sequence will not be increased in $(n-i_t)$th
step for $t=m, (m-1), \cdots, 1$ when we implement {\bf Algorithm
2.1} and will be increased in all other steps.

First, one can observe that there exist at least 2 vertices, their
distance is $p^{i_m}$.
 As the length of $A_i$ is reduced proportionally by $1/p$ in each iteration, in the $(n-i_m)$th step,
 the length of $A_i$ is $p^{i_m}$, so the two vertices  must be  in two different $A_{i_1}$ and $A_{i_2}$.
  If  ${A}_0={A}_1=\cdots= {A}_{p-1}$
is  not true, then we consider  $a\leftarrow A_0+A_1+\cdots+
A_{p-1}, $ the two vertices  must be removed, which contradicts to
Definition 3.2 requiring  there should be no decrease of nonzero
elements in operation $a\leftarrow A_0+A_1+\cdots+ A_{p-1}$ for a
hypercube.
  Thus  one
period of the sequence must be divided into $p$ equal parts in this
step, then the
 linear complexity should not be increased by $(p-1)p^{i_m}$.
 Iteratively, there exist at least 2 vertices, their distance is  $p^{i_2}$.
 In the $(n-i_2)$th step, the length of $A_i$ is $p^{i_2}$, one period of the sequence will be divided into $p$ equal parts, then the
 linear complexity should not be increased in the  $(n-i_2)$ step. Similarly, there exist at least 2 vertices, their distance is  $p^{i_1}$.
 In the $(n-i_1)$th step, the length of $A_i$ is $p^{i_1}$, one period of the sequence will be divided into $p$ equal parts in this step, then the
 linear complexity will not be increased in this step.

Second, suppose that in the $(n-i_0)$th step, one period of the
sequence is divided into $p$ equal parts, but $i_0$ is not in
$\{i_1,i_2,\cdots, i_m \}$. Then the sequence $s$ should have at
least $p^{m+1}$ vertices, which contradict to the definition of an
$m$-hypercube, which has only $p^{m}$ vertices.

Therefore, $L(s)=1+(p-1)+(p-1)p+(p-1)p^2+\cdots+(p-1)p^{n-1}-(p-1)(p^{i_1}+p^{i_2}+\cdots+p^{i_m})=1-1+p^n-(p-1)(p^{i_1}+p^{i_2}+\cdots+p^{i_m})$.

ii) As the vertex of hypercube is not a nonzero element and the
length of the vertex is 0, from {\bf Algorithm 2.1},
 the all possible linear complexity increases are:
  $$(p-1)p^{n-1}, (p-1)p^{n-2}, \cdots,(p-1), 0.$$

 Similarly, $L(s)=0-1+p^n-(p-1)(p^{i_1}+p^{i_2}+\cdots+p^{i_m})$.

 iii) As the vertex of hypercube is not a nonzero element and the length of the vertex is $j>0$, from {\bf Algorithm 2.1},
 the all possible linear complexity increases are:
  $$(p-1)p^{n-1}, (p-1)p^{n-2}, \cdots,(p-1)p^j, 0.$$

   Similarly reasoning as above, we obtain $L(s)=(1-p)(p^0+p^1+\cdots+ p^{j-1})-1+p^n-(p-1)(p^{i_1}+p^{i_2}+\cdots+p^{i_m})$.

The proof is complete now.
\end{proof}\

\noindent {\bf Example 3.1} Let $n=3, p=3$, $s^{(n)}=\{110\ 110\ 110\ 110\ 110\ 110\ 110\ 110\ 110\}$.  $s^{(n)}$ is a 2-hypercube.
Lengths of  edges are $ 3, 3^2$  respectively. The vertex of hypercube is $\{110\}$, not a nonzero element. The length of the vertex is 0. So, $\epsilon=0$, $L(s^{(n)})=0-1+3^3-(3-1)(3+3^2)=2$.

\noindent {\bf Example 3.2} Let $n=3, p=3$, $s^{(n)}=\{000\ 100\ 100\ 000\ 100\ 100\ 000\ 100\ 100\}$.  $s^{(n)}$ is a 1-hypercube.
The length of the edge is $3^2$. The vertex of hypercube is $\{000\ 100\ 100\}$, not a nonzero element. The length of the vertex is 1. So, $\epsilon=-2$,  $L(s^{(n)})=-2-1+3^3-(3-1)3^2=6$.

From Theorem   3.1, it is easy to have the following result about
the possible minimum linear complexity decrease between two
hypercubes.

\noindent {\bf Remark   3.1} Suppose that $s$ is a binary sequence
with period $p^n$, and  further $s$ is an $m$-hypercube with lengths
of edges being $ p^{i_1}, p^{i_2},\cdots ,p^{i_m}$ $(0\le i_1<
i_2<\cdots<i_m<n )$ respectively, and
$L(s)=\epsilon-1+p^n-(p-1)(p^{i_1}+p^{i_2}+\cdots+p^{i_m})$. i) If
$\epsilon=1$, and let $i_0=\min(\{0,1,2, \cdots, n-1\}-\{i_1, i_2,
\cdots, i_m\})$; ii) If $\epsilon=0$, and let $i_0=\min(\{1,2,
\cdots, n-1\}-\{i_1, i_2, \cdots, i_m\})$; iii) If
$\epsilon=(1-p)(p^0+p^1+\cdots+ p^{j-1})$, and  let
$i_0=\min(\{j+1,j+2, \cdots, n-1\}-\{i_1, i_2, \cdots, i_m\})$. Then
the maximum linear complexity less than $L(s)$ achieved by another
hypercube is
$\epsilon-1+p^n-(p-1)(p^{i_0}+p^{i_1}+p^{i_2}+\cdots+p^{i_m})$,
which is achieved by an $(m+1)$-hypercube with lengths of edges
being $p^{i_0}, p^{i_1}, p^{i_2},\cdots ,p^{i_m}$. \

Based on {\bf Algorithm 2.1}, it is easy to give the following
result about  standard hypercube decomposition.

\noindent {\bf Theorem   3.2} Suppose that $s$ is a binary sequence
with period $p^n$, and $L(s)=\epsilon-1+p^n-(p-1)(p^{i_1}+p^{i_2}+\cdots+p^{i_m})$, $\epsilon\in\{0,1\}$,
where $0\le i_1< i_2<\cdots<i_m<n$, then the sequence $s$ can be
decomposed into several  hypercubes, $s=h_1\bigoplus h_2\bigoplus h_3\bigoplus \cdots$,
and only  hypercube $h_1$ has the
linear complexity $L(s)$, other
hypercubes possess distinct linear complexities  less than
$L(s)$.

In summary, one can see that any sequence $s$ can be decomposed into
 the sum of several hypercubes with descending order of linear complexity.
 As the decomposition is not unique, it may be hard to investigate the
 possible decompositions further and we will tackle this problem in future. Luckily, the first hypercube has
 the same linear complexity with the original sequence. We will
 investigate $k$-error linear complexity and other important issues by using this decomposition in next section.

\section{$k$-error Linear Complexity and Stability of $p^n$ Periodic Binary Sequences}

In this section, we will investigate two important problems. One is
the minimum value $k$ for which the $k$-error linear complexity of
$s$ is strictly less than the linear complexity $L(s)$ of $s$. This
problem has attracted much attention recently for a $p^n$-periodic
binary sequence $s$. Currently the best result is given by Meildl
\cite{Meidl} by using polynomial approach and only an upper bound is
obtained. By using the hypercube decomposition, we solve this
problem completely in this section. Consequently, the stability of a
$p^n$-periodic binary sequence $s$ is investigated. Second, as the
importance of hypercube, we give a numeration formula of all
possible hypercubes with a given linear complexity with partial
characterization of construction. This will pave a path for its
further applications in future.

\subsection{ $k$-error linear
complexity  of a $p^n$-periodic  binary sequence}

Based on hypercube theory, linear
complexity is discussed in previous section. Next we will investigate the $k$-error linear
complexity  for a $p^n$-periodic  binary sequence $s$.

For a $2^n$-periodic binary sequence $s$, one important result, proved by Kurosawa et al.  \cite{Kurosawa}, is
that the minimum value $k$ denoted as $m(s)$   for which the $k$-error linear
complexity of  $s$ is strictly less
than the linear complexity $L(s)$ of $s$ is determined by
$m(s)=2^{W_H(2^n-L(s))}$, where $W_H(a)$ denotes the Hamming
weight of the binary representation of an integer $a$.

For a $p^n$-periodic binary sequence $s$, with a polynomial  approach, Meidl \cite{Meidl}
studied the minimum value  $k$  for which the $k$-error
linear complexity is strictly less than the linear complexity of a $p^n$-periodic  binary sequence $s$,
where $p$ is an odd prime and 2 is a primitive root modulo $p^2$. The following upper bound on $m(s)$ is established in \cite{Meidl}.
$$m(s)\le (\frac{p-1}{2})^\delta p^{W_H(p^n-L(s))}$$ where $\delta=(\epsilon+1)\mod 2,
\epsilon\in\{0,1\}$. This is the best known result in literature.

With the proposed hypercube theory, we further study  $m(s)$  for a $p^n$-periodic  binary sequence $s$.
 We first consider sequences with only one hypercube.

\noindent {\bf Theorem   4.1} Suppose that $s$ is a  hypercube with
period $p^n$, and
$L(s)=\epsilon-1+p^n-(p-1)(p^{i_1}+p^{i_2}+\cdots+p^{i_m})$, where
$\epsilon\in\{0,1, (1-p)(p^0+p^1+\cdots+ p^{q-1})\}$, in which $q$
is the length of the vertex, and
 $0\le i_1< i_2<\cdots<i_m<n$.  Then

 i) If the vertex is a nonzero element, then
$m(s)=\begin{array}{l} p^m.
\end{array}$

ii) If the vertex is  a  $p$-tuple $\{A_0,A_1,\cdots, A_{p-1}\}$ and
 the length of the vertex is 0.  Assume there are $l$ nonzero elements in the $p$-tuple, then

$$m(s)=\left\{\begin{array}{l}
lp^m, \ \ \ \ \ \ \ \ \ \ \   \ l<p/2\ \   \\
(p-l)p^m, \ \ \  \ \ \mbox{otherwise}
\end{array}\right.$$

iii) If the vertex is  a  $p$-tuple $\{A_0,A_1,\cdots, A_{p-1}\}$
and the vertex has a length $q>0$. Now assume there are $l$ nonzero
elements in the tuple and the  vertex $p$-tuple $\{A_0,A_1,\cdots,
A_{p-1}\}$ can be changed to a nonzero $p$-tuple
$\{\tilde{A_0},\tilde{A_1},\cdots, \tilde{A_{p-1}}\}$ with $j$
elements change, such that $\tilde{A_0}=\tilde{A_1}=\cdots=
\tilde{A_{p-1}}$ (refer to Appendix iV) on how to calculate $j$),
then

 $$m(s)=\left\{\begin{array}{l}
lp^m, \ \ \ \ \ \ \ \ \ \ \    l<j\ \   \\
jp^m, \ \ \  \ \ \ \ \ \ \ j<l
\end{array}\right.$$

\noindent \begin{proof}\ We prove the three cases separately as
follows.

i)  If the vertex is a nonzero element. Based on {\bf Algorithm
2.1}, to decrease the linear complexity  of $s$, there are two
possibilities: to remove or to add some nonzero elements.

Suppose in the $k_0$th step of {\bf Algorithm 2.1}, $1\le k_0\le n$,
$A_0=A_1=\cdots= A_{p-1}$ are not true at first time. Then  the
 linear complexity of $s$ is increased by $(p-1)p^{n-k_0}$.  In order to avoid being increased by $(p-1)p^{n-k_0}$ for the changed sequence, we may change $s$, so that
$A_0=A_1=\cdots= A_{p-1}$ for the changed sequence.
 According {\bf Remark 2.1}, we note that after the $k_0$th step, the sum of all possible increase for linear complexity of the changed sequence is $p^{n-k_0}$.
 Thus the change of the $k_0$th step will lead to the  decrease of final linear complexity.

One option to change $s$ is  to remove all nonzero elements in
$\{A_0, A_1, \cdots,  A_{p-1} \}$. As $s$ is a hypercube, this is
equivalent to remove the hypercube $s$ with $p^m$ nonzero elements.

Another option is to  add  some nonzero elements and delete some
other nonzero elements (if $k_0=n$, we can only add nonzero elements
as only one nonzero element left in this case), such that
$A_0=A_1=\cdots= A_{p-1}$ after changes. From Definition 3.2, there
should be no decrease of nonzero elements in operation $a\leftarrow
A_0+A_1+\cdots+ A_{p-1}$ for hypercube $s$.

Assume

 $$A_i=\left(\begin{array}{c}a_{1i}\\ a_{2i}\\ \vdots\\
a_{p^{n-k_0}i}\end{array}\right), 0\le i<p.
A=\left(\begin{array}{c}a_{10}\ \ a_{11}\ \ \ \ \ \ \cdots\ \ a_{1,p-1}\\
a_{20}\ \ a_{21}\ \ \ \ \ \ \cdots\ \ a_{2,p-1}\\
 \ \ \ \ \vdots\\
a_{p^{n-k_0},0}\ a_{p^{n-k_0},1}\ \cdots\
a_{p^{n-k_0},p-1}\end{array}\right).$$

One can derive that there is at most one nonzero element in each row
of $A$. We can further assume there are $p^x$ nonzero elements in
$A$. In this case, there are $p^m-p^y$ elements removed in previous
operation and $x+y=m$. In order to make changes in $A_i$ such that
$A_0=A_1=\cdots= A_{p-1}$, we must change the nonzero rows to all
zeros or all ones. Now if $a_{ji}=1$ $ 0\le i <p$, we change the $j$
row to all ones and  all other rows be zero such that
$A_0=A_1=\cdots= A_{p-1}$. The number of elements changed is
$$p^y(p-1)+p^y(p^x-1)=p^m+p^y(p-2)>p^m$$

In above formula, $p^y(p-1)$ represents the changes corresponding to
the $j$ row in original sequence and $p^y(p^x-1)$ represents the
changes of all other nonzero changes in the original sequence.  For
general case, let $t$ rows be all nonzero elements. Then The number
of elements changed is
$$tp^y(p-1)+p^y(p^x-t)=p^m+tp^y(p-2)>p^m$$

In summary, the minimal change is to remove all the nonzero
elements, thus $m(s)=p^m$.

ii) If the vertex is  a  $p$-tuple $\{A_0,A_1,\cdots, A_{p-1}\}$
with $l$ nonzero  elements, the length of the vertex is 0. In this
case, we consider the $n$th step of {\bf Algorithm 2.1} for sequence
$s$, $\{A_0, A_1, \cdots, A_{p-1} \}$ is  a vertex. As $A_0+ A_1+
\cdots+ A_{p-1} =\{0,0.\cdots,0\}$, $A_0=A_1=\cdots= A_{p-1}$ are
not true. From Algorithm 2.1,  the
 linear complexity is increased by $(p-1)p^{0}$. To avoid increasing by $(p-1)p^{0}$, we may change $s$, so that $A_0=A_1=\cdots= A_{p-1}$.


We have to remove all nonzero elements in $\{A_0, A_1, \cdots,  A_{p-1} \}$, which means to remove the hypercube $s$ with $lp^m$ nonzero elements, or add  some nonzero elements, such that
$A_0=A_1=\cdots= A_{p-1}$, which means to add $(p-l)\times p^m$ nonzero elements to the hypercube $s$.

Assume in the $k_0$th step of Algorithm 2.1, $1\le k_0< n$,
$A_0=A_1=\cdots= A_{p-1}$ are not true. Similar to the analysis of
i), we have,
$$m(s)=\left\{\begin{array}{l}
lp^m, \ \ \ \ \ \ \ \ \ \ \   \ l<p/2\ \   \\
(p-l)p^m, \ \ \  \ \ \mbox{otherwise}
\end{array}\right.$$


iii)If the vertex has length $q>0$, and is  a  $p$-tuple
$\{A_0,A_1,\cdots, A_{p-1}\}$ with $l$ nonzero  elements,  and the
vertex can be changed to a nonzero $p$-tuple
$\{\tilde{A_0},\tilde{A_1},\cdots, \tilde{A_{p-1}}\}$ with the least
$j$ elements change, such that $\tilde{A_0}=\tilde{A_1}=\cdots=
\tilde{A}_{p-1}$.


Now we consider the $(n-q)$th step of {\bf Algorithm 2.1} for
sequence $s$, $\{A_0, A_1, \cdots, A_{p-1} \}$ is  a vertex. As
$A_0+ A_1+ \cdots+ A_{p-1} =\{0,0.\cdots,0\}$, $A_0=A_1=\cdots=
A_{p-1}$ are not true. From Algorithm 2.1,  the
 linear complexity is increased by $(p-1)p^{q}$. To avoid being increased by $(p-1)p^{n-q}$ for the changed sequence, we need to change $s$, so that $A_0=A_1=\cdots= A_{p-1}$.

As $\{A_0, A_1, \cdots,  A_{p-1} \}$ is  a vertex. If $l<j$, we have to remove all the nonzero elements in hypercube $s$ to  decrease the linear complexity. 

Otherwise, if $l>j$, with at least $j$ elements change, the  vertex can be changed to a nonzero $p$-tuple $\{\tilde{A_0},\tilde{A_1},\cdots, \tilde{A_{p-1}}\}$, such that $\tilde{A_0}=\tilde{A_1}=\cdots= \tilde{A_{p-1}}$,  which decreases  the linear complexity. 

Assume in the $k_0$th step of Algorithm 2.1, $1\le k_0< n-q$,
$A_0=A_1=\cdots= A_{p-1}$ are not true. Similar to the analysis of
i), we have,

$$m(s)=\left\{\begin{array}{l}
lp^m, \ \ \ \ \ \ \ \ \ \ \    l<j\ \   \\
jp^m, \ \ \  \ \ \ \ \ \ \ j<l
\end{array}\right.$$

\end{proof}\

It should be noted that the number of $j$ in Appendix iv) depends on
the structure of the vertex and it is hard to give its direct
relation with the sequence $s$. Anyway, we can compute this value
easily with a given hypercube.

The following examples are given to illustrate Theorem 4.1.

Let $n=3, p=3$,  $s^{(n)}=\{110\ 000\ 000\ 110\ 000\ 000\ 110\ 000\ 000\}.$  As    hypercube $s^{(n)}$ has $2\times3$ nonzero elements and $2>3/2$, thus $m(s^{(n)})=(3-2)3=3$. 

Let $n=2, p=5$, $s^{(n)}=\{11110\ 11110\ 11110\ 11110\ 11110\}$. Then $L(s^{(n)})=-1+5^2-4\times5$. As    hypercube $s^{(n)}$ has $4\times5$ nonzero elements and $4>5/2$, thus $m(s^{(n)})=(5-4)5=5$. 

 Let $n=3, p=3$, $s^{(n)}=\{000\ 100\ 100\ 000\ 100\ 100\ 000\ 100\ 100\}$.  $s^{(n)}$ is a 1-hypercube.
  $L(s^{(n)})=-1+3^3-(3-1)(1+3^2)=6$. As    hypercube $s^{(n)}$ has $2\times3$ nonzero elements and $2>3/2$, thus $m(s^{(n)})=(3-2)3=3$.

 For a $p^n$-periodic binary sequence $s$, Meidl \cite{Meidl} obtained sharp lower and upper
bounds on $m(s)$.
 By Theorem 1 in \cite{Meidl}, for $s^{(n)}=\{11110\ 11110\ 11110\ 11110\ 11110\}$, $m(s^{(n)})\le  \frac{5-1}{2} \times5^1=10$,  which is  greater than $m(s^{(n)})=5$.

By Theorem 1 in \cite{Meidl}, for $s^{(n)}=\{000\ 100\ 100\ 000\ 100\ 100\ 000\ 100\ 100\}$, $m(s^{(n)})\le  \frac{3-1}{2} \times3^2=9$, which is  greater than $m(s^{(n)})=3$.

So the result obtained here is  much more precise.

\

 For a general binary sequence $s$ consisting of different hypercubes, $s=h_1\bigoplus h_2\bigoplus h_3\bigoplus \cdots$, the following theorem establishes a relationship between the greatest hypercube $h_1$ of $s$ and $m(s)$.

 \noindent {\bf Theorem   4.2} Suppose that $s$ is a binary sequence
with period $p^n$, and
$L(s)=\epsilon-1+p^n-(p-1)(p^{i_1}+p^{i_2}+\cdots+p^{i_m})$, where
$\epsilon\in\{0,1, (1-p)(p^0+p^1+\cdots+ p^{q-1})\}$ and
 $0\le i_1< i_2<\cdots<i_m<n$, and $h$ is a hypercube with linear complexity $L(s)$ in the standard hypercube decomposition of $s$.

 i) If the vertex of $h$ is a nonzero element, then
$m(s)=\begin{array}{l} p^m
\end{array}$

ii) If the vertex of $h$  has length 0, and   is  a  $p$-tuple
$\{A_0,A_1,\cdots, A_{p-1}\}$ with $l$ nonzero  elements,   then

$$m(s)=\left\{\begin{array}{l}
lp^m, \ \ \ \ \ \ \ \ \ \ \   \ l<p/2\ \   \\
(p-l)p^m, \ \ \  \ \ \mbox{otherwise}
\end{array}\right.$$

iii) If the vertex of $h$ has length $q>0$, and is  a  $p$-tuple
$\{A_0,A_1,\cdots, A_{p-1}\}$ with $l$ nonzero  elements,  and the
vertex can be changed to a nonzero $p$-tuple
$\{\tilde{A_0},\tilde{A_1},\cdots, \tilde{A_{p-1}}\}$ with at least
$j$ elements change, such that $\tilde{A_0}=\tilde{A_1}=\cdots=
\tilde{A}_{p-1}$,    then

$$m(s)=\left\{\begin{array}{l}
lp^m, \ \ \ \ \ \ \ \ \ \ \    l<j\ \   \\
jp^m, \ \ \  \ \ \ \ \ \ \ j<l
\end{array}\right.$$


\begin{proof}\
To  decrease the linear complexity  of $s$, we only need to consider
hypercube $h$   with linear complexity $L(s)$. Based on Theorem
4.1, the result is obvious.
\end{proof}\

\

Now we consider an application of Theorem 4.2.
It is known that both high
linear complexity  and high $k$-error
linear complexity are necessary for the security of a key stream. Now we give a class of sequences $s$ with the largest $k$-error
linear complexity and $L(s)=L_k(s)$.


Let $s$ be the binary sequence
$\{\overbrace{11\cdots11}^{p^k}0\cdots0\}$. Its period is $p^n$, and
there are only $p^k$ continuous nonzero elements at the beginning of
the sequence.  Then it is a $k$-hypercube with vertex being a
nonzero element 1. By Theorem 4.2, $m(s)=p^k$. So, after at most
$e(0\le e\le p^k-1)$ elements change in a period of the above
sequence are changed, the linear complexity of all new sequences are
not decreased, thus the original sequence possesses stable $e$-error
linear complexity.
 The
$p^{k-1},\cdots, (p^k-2)$ or $(p^k-1)$-error linear complexity of $s$ are
all $p^n-(p^k-1)$.

%

So  we have
the following important corollary.

\noindent {\bf Corollary   4.1} For $ p^{l-1}\le k<p^{l}$, we can construct one $p^n$-periodic  binary sequence  $s$
with stable $k$-linear complexity $p^n-(p^l-1)$, such that
$$L_k(s)=\max\limits_tL_k(t)$$ where   $t$ is any
$p^n$-periodic  binary sequence.

It is worthy to mention that there are $(3^{p^k})^{n-k}$ sequences
with linear complexity $p^n-(p^k-1)$ derived from sequence
$\{\overbrace{11\cdots11}^{p^k}0\cdots0\}$. For example, let $n=2,
p=3$. From $s^{(n)}=\{111\ 000\ 000\}$, we have the following $3^3$
sequences with linear complexity $3^2-(3^1-1)=7$.

$\{111\ 000\ 000\}$, \{011\ 100\ 000\}$, \{011\ 000\ 100\}$

$\{101\ 010\ 000\}$, \{001\ 110\ 000\}$, \{001\ 010\ 100\}$

$\{101\ 000\ 010\}$, \{001\ 100\ 010\}$, \{001\ 000\ 110\}$

$\cdots \cdots$

\

It is reminded that the CELCS (critical error linear complexity
spectrum) has been studied by Etzion et al. \cite{Etzion}. The CELCS
of a sequence $s$ consists of the ordered set of points $(k,L_k(s))$
satisfying $L_k(s)> L_{k'}(s)$, for $k'>k$; these are the points
where a decrease occurs in the $k$-error linear complexity, and thus
they are called critical points.


In fact, $m(s)$ is the first  critical point, next we consider the second critical point.
We define $m_1(s)$ as the minimum  $k$ for which the $k$-error
linear complexity is strictly less than $L_{m(s)}(s)$.
We first consider sequences with only one hypercube.

\noindent {\bf Proposition   4.1} Suppose that $s$ is a binary
sequence with period $p^n$, and
$L(s)=\epsilon-1+p^n-(p-1)(p^{i_1}+p^{i_2}+\cdots+p^{i_m})$, where
$\epsilon\in\{0,1, (1-p)(p^0+p^1+\cdots+ p^{q-1})\}$ and
 $0\le i_1< i_2<\cdots<i_m<n$, and $s$ is a hypercube.

 i) If the vertex is a nonzero element, then there is no second critical point.

 ii) Assume that the vertex of $h$  has length 0, and   is  a  $p$-tuple $\{A_0,A_1,\cdots, A_{p-1}\}$ with $l$ nonzero  elements. If $p/2<l<p$ then
 $m_1(s)=lp^m$. Otherwise, there is no second critical point.

iii) Assume that the vertex of $h$ has length $q>0$, and is  a
$p$-tuple $\{A_0,A_1,\cdots, A_{p-1}\}$ with $l$ nonzero  elements,
and the  vertex can be changed to a nonzero $p$-tuple
$\{\tilde{A_0},\tilde{A_1},\cdots, \tilde{A_{p-1}}\}$ with at least
$j$ elements change, such that $\tilde{A_0}=\tilde{A_1}=\cdots=
\tilde{A_{p-1}}$.
 If $j<l$ then
 $m_1(s)=lp^m$. Otherwise, there is no second critical point.
\begin{proof}\ i) It is obvious from Theorem   4.1.

ii) If $p/2<l<p$, from Theorem   4.1, $m(s)=(p-l)p^m$. In this case,
in order to further decrease  the linear complexity  of $s$, we have
to remove the hypercube $s$. So $m_1(s)=lp^m$. Otherwise, there is
no second critical point.

iii) If $l>j$, from Theorem   4.1, $m(s)=jp^m$. In this case, in
order to further decrease  the linear complexity  of $s$, we have to
remove the hypercube $s$. So $m_1(s)=lp^m$. Otherwise, there is no
second critical point.

\end{proof}\

 Though this problem is completely solved for a hypercube, for a general binary sequence consisting of different hypercubes, the following example is
 presented to illustrate the difficulty to compute $m_1(s)$.
 In fact, $m(s)$ is only related to the largest hypercube of $s$, but $m_1(s)$ may be related to all possible hypercubes of $s$.

Let $n=3, p=3$,  $s^{(n)}=\{110\ 000\ 000\ 111\ 000\ 000\ 111\ 000\
000\}.$ Thus $s^{(n)}$ consists of hypercube $\{110\ 000\ 000\ 000\
000\ 000\ 000\ 000\ 000\}$ and     hypercube $\{000\ 000\ 000\ 111\
000\ 000\ 111\ 000\ 000\}$

From Theorem 4.2, $m(s^{(n)})=(3-2)=1$.  As $\{110\ 000\ 000\ 000\
000\ 000\ 000\ 000\ 000\}$ becomes $\{111\ 000\ 000\ 000\ 000\ 000\
000\ 000\ 000\}$, so the new $s^{(n)}$ is $\{111\ 000\ 000\ 111\
000\ 000\ 111\ 000\ 000\}$, which is a 2-hypercube. Therefore  to
further decrease  the linear complexity, we have to remove the
hypercube. So $m_1(s^{(n)})=3\times3-1=8$.

%
%

\

\subsection{Counting of $p^{n}$-periodic binary sequences of given linear complexity with one  hypercube}

Next we consider the number  of sequences with exactly one  hypercube by a construction approach.
Suppose that $s$ is a binary sequence with period  $p^n$, and
$L(s)=\epsilon-1+p^n-(p-1)(p^{i_1}+p^{i_2}+\cdots+p^{i_m})$,
where  $0\le i_1<
i_2<\cdots<i_m<n$. We first derive the counting formula of
$m$-hypercubes with the same linear complexity.

\noindent {\bf Theorem  4.3}  Suppose that $s$ is a binary sequence
with period  $p^n$,  $L(s)=\epsilon-1+p^n-(p-1)(p^{i_1}+p^{i_2}+\cdots+p^{i_m})$,  and
 $0\le i_1< i_2<\cdots<i_m<n$. Let
$$C=p^{p^m n-(p^m-p^{m-1})i_{m}-\cdots-(p^2-p)i_2-(p-1)i_1-\frac{p^{m+1}-p}{p-1}}$$

If $\epsilon=1$, then the vertex is a nonzero element, and the number of all $m$-hypercubes $e$ with $L(e)=L(s)$ is $C$.

If $\epsilon=0$, and  the vertex is with length 0 and
has $l$ nonzero elements, then the number of all $m$-hypercubes $e$ with $L(e)=L(s)$ is $\left(\begin{array}{c}p\\l\end{array}\right)\left(\frac{C}{p}\right)^l,  1<l<p$.

\begin{proof}\
We first consider the case of $\epsilon=1$.

Suppose that $s^{(i_1)}$ is a $p^{i_1}$-periodic binary sequence with linear complexity  $p^{i_1}$ and $W_H(s^{(i_1)})=1$,
then  the number of these $s^{(i_1)}$ is $p^{i_1}$

So the number of $p^{i_1+1}$-periodic binary sequences $s^{(i_1+1)}$ with linear complexity $p^{i_1+1}-(p-1)p^{i_1}=p^{i_1}$ and $W_H(s^{(i_1+1)})=p$ is also $p^{i_1}$.

For $i_2>i_1$,
if $p^{i_2}$-periodic binary sequences $s^{i_2}$ with linear complexity $p^{i_2}-(p-1)p^{i_1}$ and $W_H(s^{(i_2)})=p$,
then $p^{i_2}-(p-1)p^{i_1}-(p^{i_1+1}-(p-1)p^{i_1})=(p-1)p^{i_2-1}+(p-1)p^{i_2-2}+\cdots+(p-1)p^{i_1+1}$.

 Based on {\bf Algorithm 2.1},
the number of these $s^{i_2}$ can be given by
$(p^p)^{i_2-i_1-1}\times p^{i_1}=p^{pi_2-(p-1)i_1-p}$.

(The following examples are given to illustrate the proof.

Suppose that $i_1=1, i_2=3, p=3$, then  $(p^p)^{i_2-i_1-1}=27$ sequences

\{100100100\ 000000000\ 000000000\},

\{100100000\ 000000100\ 000000000\},

\{100100000\ 000000000\ 000000100\},

\{100000100\ 000100000\ 000000000\},

\{100000000\ 000100100\ 000000000\},

\{100000000\ 000100000\ 000000100\},

\{100000100\ 000000000\ 000100000\},

\{100000000\ 000000100\ 000100000\},

\{100000000\ 000000000\ 000100100\},

$\cdots \cdots $

of $s^{(i_2)}$
correspond to a sequence \{100100100\} of $s^{(i_1+1)}$.
)

So the number of $p^{i_2+1}$-periodic binary sequences $s^{(i_2+1)}$ with linear complexity $p^{i_2+1}-(p-1)(p^{i_2}+p^{i_1})=p^{i_2}-(p-1)p^{i_1}$ and $W_H(s^{(i_2+1)})=p^2$ is also $p^{pi_2-(p-1)i_1-p}$.

For $i_3>i_2$, based on {\bf Algorithm 2.1}, if $p^{i_3}$-periodic
binary sequences $s^{i_3}$ with linear complexity
$p^{i_3}-(p-1)(p^{i_2}+p^{i_1})$ and $W_H(s^{(i_3)})=p^2$, then the
number of these $s^{i_3}$ can be given by
$(p^{p^2})^{i_3-i_2-1}\times
p^{pi_2-(p-1)i_1-p}=p^{p^2i_3-(p^2-p)i_2-(p-1)i_1-p-p^2}$.

$\cdots\cdots$

So the number of $p^{i_m+1}$-periodic binary sequences $s^{(i_m+1)}$ with linear complexity
$p^{i_m+1}-(p-1)(p^{i_1}+p^{i_2}+\cdots+p^{i_m})=p^{i_m}-(p-1)(p^{i_1}+p^{i_2}+\cdots+p^{i_{m-1}})$
and $W_H(s^{(i_m+1)})=p^m$ is also $$p^{p^{m-1}i_{m}-\cdots-(p^2-p)i_2-(p-1)i_1-p-p^2-\cdots-p^{m-1}}$$

For $n>i_m$,
if $p^{n}$-periodic binary sequences $s^{(n)}$ with linear complexity $p^n-(p-1)(p^{i_1}+p^{i_2}+\cdots+p^{i_m})$ and $W_H(s^{(n)})=p^m$,
then the number of these $s^{(n)}$ can be given by
{
\scriptsize
\begin{eqnarray*}&&(p^{p^m})^{n-i_m-1}\times p^{p^{m-1}i_{m}-\cdots-(p^2-p)i_2-(p-1)i_1-p-\cdots-p^{m-1}}\\
&=&p^{p^m n-(p^m-p^{m-1})i_{m}-\cdots-(p^2-p)i_2-(p-1)i_1-p-\cdots-p^{m-1}-p^m}\\
&=&p^{p^m n-(p^m-p^{m-1})i_{m}-\cdots-(p^2-p)i_2-(p-1)i_1-\frac{p^{m+1}-p}{p-1}}
\end{eqnarray*}
}

This completes the proof of the first part.

Now consider the case of $\epsilon=0$.

Suppose that $s^{(i_1)}$ is a $p^{i_1}$-periodic binary sequence with linear complexity  $p^{i_1}-1$ and $W_H(s^{(i_1)})=l$,
then  the number of these $s^{(i_1)}$ is $\left(\begin{array}{c}p\\l\end{array}\right)p^{l(i_1-1)}$

So the number of $p^{i_1+1}$-periodic binary sequences $s^{(i_1+1)}$ with linear complexity $p^{i_1+1}-(p-1)p^{i_1}-1=p^{i_1}-1$ and $W_H(s^{(i_1+1)})=lp$ is also $\left(\begin{array}{c}p\\l\end{array}\right)p^{l(i_1-1)}$.

$\cdots\cdots$

Similarly, we have the following result.

If $p^{n}$-periodic binary sequences $s^{(n)}$ with linear complexity $p^n-(p-1)(p^{i_1}+p^{i_2}+\cdots+p^{i_m})-1$,  the vertex  is with length 0 and $W_H(s^{(n)})=lp^m$,
then the number of these $s^{(n)}$ can be given by
$$\left(\begin{array}{c}p\\l\end{array}\right)\left(\frac{C}{p}\right)^l.$$

\end{proof}\

For  a binary sequence
with period  $2^n$,
Etzion et al.
first proved the following Proposition 4.2 in \cite{Etzion}. We  proved it with cube theory in  \cite{Zhou_Liu2013} as well.
It is also a special case of Theorem 4.3.

\noindent {\bf Proposition  4.2}  Suppose that $s$ is a binary sequence
with period  $2^n$, and $L(s)=2^n-(2^{i_1}+2^{i_2}+\cdots+2^{i_m})$,
where  $0\le i_1< i_2<\cdots<i_m<n$. If sequence $e$ is an $m$-cube with $L(e)=L(s)$, then the number of  sequence $e$ is
$$2^{2^{m}n-2^{m-1}i_m-\cdots-2i_2-i_1-2^{m+1}+2}$$

\


From Definition 3.3, we know that there are three types of vertices.
They have $1, l$( the vertex  is with length 0) or  $lj$( the vertex
is with nonzero length) elements respectively, $0<l<p$. In  Theorem
4.3, we investigate a $p^n$-periodic binary sequence with vertex
having 1 or $l$ nonzero elements.   The counting formula of
$m$-hypercubes with  vertex having nonzero length  remains to be
solved in future.

\

\section{Conclusions}

For a $p^n$-periodic  binary sequence, where $p$ is an odd prime and 2 is a primitive root modulo $p^2$,
by studying  sequences with minimum Hamming weight, a new tool called hypercube theory has been
developed.
A general hypercube decomposition approach has been given. Also,  a characterization has been presented about
 the first decrease  in the
$k$-error linear complexity for a $p^n$-periodic binary sequence $s$ based on the proposed hypercube theory.
One very important application is to construct sequences with
the maximum stable $k$-error linear complexity.
 Finally, a counting
formula for $m$-hypercubes with the same linear complexity has been derived.

 The hypercube structure of a $p^n$-periodic binary sequence is closely related to its linear complexity and
$k$-error linear complexity. So it is
 is very important in
investigating critical error linear complexity spectrum proposed by Etzion et al, which is our future
work.

 \section*{ Acknowledgment}
 The research was partially supported by
Anhui Natural Science Foundation(No.1208085MF106).

\appendix

\section{How I became inspired}
1). The construction of $\tilde{A}_i$  in Step 3 of {\bf Algorithm
3.1}

If ${A}_0={A}_1=\cdots= {A}_{p-1}$
is  not true, then
 $a\leftarrow A_0+A_1+\cdots+ A_{p-1}, $
  we now consider $  A_0+A_1+\cdots+ A_{p-1}$.

If the number of nonzero elements in $a$ is less than the sum of the
number of nonzero  elements in each $A_i$ for $0\le i<p$, then we
can change the nonzero elements in $A_0, A_1,\cdots, A_{p-1}$
accordingly  such that

i)  $ \tilde{A}_0+\tilde{A}_1+\cdots+ \tilde{A}_{p-1} $ is still
equal to the original $a$,

ii)
  the number of nonzero elements in $a$ is the same
as the sum of the number of nonzero elements in $\tilde{A}_i$ for
$0\le i<p$,

iii) $\tilde{A}_0=\tilde{A}_1=\cdots= \tilde{A}_{p-1}$
is still not true.

In fact, we can assume that

 $A_i=\left(\begin{array}{c}a_{1i}\\ a_{2i}\\ \vdots\\
a_{qi}\end{array}\right), 0\le i<p$. Then
$a=\left(\begin{array}{c}a_{10}+a_{11}+\cdots+a_{1,p-1}\\
a_{20}+a_{21}+\cdots+a_{2,p-1}\\ \vdots\\
a_{q0}+a_{q1}+\cdots+a_{q,p-1}\end{array}\right).$


Suppose that for some $j$, we have
$a_{j0}+a_{j1}+\cdots+a_{j,p-1}=1$. Then we can keep the first
nonzero element in $\{ a_{j0}, a_{j1}, \cdots, a_{j,p-1}\}$
unchanged, let all other nonzero  elements be zero.

Suppose that $a_{j0}+a_{j1}+\cdots+a_{j,p-1}=0$.  Then change all
nonzero elements in $\{ a_{j0}, a_{j1}, \cdots, a_{j,p-1}\}$ into
zero.

After above changes, it is obvious that we have

i)   $ \tilde{A}_0+\tilde{A}_1+\cdots+ \tilde{A}_{p-1} $ is still
equal to the original $a$,

ii)   the number of nonzero elements in $a$ is the same
as the sum of the number of nonzero elements in $\tilde{A}_i$ for
$0\le i<p$.

As $\{A_0, A_1,\cdots, A_{p-1}\}$ is not vertex, thus $a$ is not a
zero vector. Suppose that $a_{j0}+a_{j1}+\cdots+a_{j,p-1}=1$. Then
there exists $i_3$ such that $a_{j,i_3}=1$. In this case,  $
\tilde{A}_0\ne\tilde{A}_{i_3}$, which implies that
$\tilde{A}_0=\tilde{A}_1=\cdots= \tilde{A}_{p-1}$ is still not true.

2). The proof of Linear complexity $h_1$ in Step 4 of {\bf Algorithm
3.1}

Step 4. Repeat  above operations, until that $a$ is
reduced to one vertex. In above process, keep all possible changes
in $s^{(n)}$ and it will finally become a hypercube $h_1$ with
linear complexity $L(s^{(n)}_0)$.

\noindent \begin{proof}\

As  the number of nonzero elements in $a$ is the same
as the sum of the number of nonzero elements in $\tilde{A}_i$ for
$0\le i<p$.
In Step 4, when we apply Algorithm 2.1 to $h_1$, thus there is no  decrease of nonzero element  in $h_1$ in the operation $a\leftarrow A_0+A_1+\cdots+ A_{p-1}, $(except for the last operation), so $h_1$ is  a hypercube.

As the following steps are still true

i) $ \tilde{A}_0+\tilde{A}_1+\cdots+ \tilde{A}_{p-1} $ is still
equal to the original $a$,

iii) $\tilde{A}_0=\tilde{A}_1=\cdots= \tilde{A}_{p-1}$
is still not true.

So if we  apply {\bf Algorithm 2.1} to $h_1$, we will have the same
process as applying  {\bf Algorithm 2.1} to $s^{(n)}_0$, so $h_1$
has linear complexity $L(s^{(n)}_0)$.

\end{proof}\

3). The proof of linear complexity of $L(h_2)$ is less than $L(s)$
in Step 5 of {\bf Algorithm 3.1}.

Step 5. With $s_0^{(n)}\bigoplus h_1$, where $s_0^{(n)}$ is the original sequence, run Step 1 to Step 4. We obtain a hypercube $h_2$ with linear complexity less than $L(s_0^{(n)})$.

\noindent \begin{proof}\

We first prove that $s_0^{(n)}\bigoplus h_1$ has linear complexity less than $L(s_0^{(n)})$.

When we apply {\bf Algorithm 3.1} to $s_0^{(n)}$, suppose that the
last time we modify the nonzero elements in $A_0, A_1,\cdots,
A_{p-1}$ in Step 3 is at the $k_0$th step,  $1\le k_0< n$.

Now let us define  $B_i=\left(\begin{array}{c}b_{1i}\\ b_{2i}\\ \vdots\\
b_{qi}\end{array}\right)$ ro represent the changes when we change  $A_i=\left(\begin{array}{c}a_{1i}\\
a_{2i}\\ \vdots\\ a_{qi}\end{array}\right),  0\le i<p$.
Specifically, if $a_{ji}$ is changed, then define $b_{ji}=1$,
otherwise let $b_{ji}=0$, $1\le j \le q$.

As  the number of nonzero  elements being changed in $\{ a_{j0},
a_{j1}, \cdots, a_{j,p-1}\}$ is either 0 or an even number, $1\le j
\le q$, thus $B_0+B_1+\cdots+ B_{p-1}=\{0,0,\cdots,0\}$.

When we apply Algorithm 2.1 to $s_0^{(n)}\bigoplus h_1$, if the
process before  the $k_0$th step is the same as applying  Algorithm
2.1 to $s^{(n)}_0$, then we will obtain $B_0, B_1,\cdots, B_{p-1}$
in Step 3  at the $k_0$th step. As $B_0+B_1+\cdots+
B_{p-1}=\{0,0,\cdots,0\}$, Algorithm 2.1 will end and no linear
complexity increase. As $A_0+A_1+\cdots+
A_{p-1}\ne\{0,0,\cdots,0\}$, {\bf Algorithm 2.1} applying to
$s^{(n)}_0$ will continue and the linear complexity will increase as
${A}_0={A}_1=\cdots= {A}_{p-1}$ is  not true. Thus
$s_0^{(n)}\bigoplus h_1$ has linear complexity less than
$L(s_0^{(n)})$ in this case.

If  one step before  the $k_0$th step is different from applying
{\bf Algorithm 3.1} to $s^{(n)}_0$, then  there exists the minmum
$k_1$, $k_1<k_0$, such that at the $k_1$th step,
 ${A}_0={A}_1=\cdots= {A}_{p-1}$
is  not true, but  ${B}_0={B}_1=\cdots= {B}_{p-1}$. In this case,
there is the first increase of linear complexity for $s^{(n)}_0$ but
no increase for $s_0^{(n)}\bigoplus h_1$. From Remark 2.1, we know
the first increase dominates the linear complexity, so
$s_0^{(n)}\bigoplus h_1$ has linear complexity less than
$L(s_0^{(n)})$.

According to the decomposition, the linear complexity of hypercube
$h_2$ is equal to $L(s_0^{(n)}\bigoplus h_1)$, thus hypercube $h_2$
has linear complexity less than $L(s_0^{(n)})$.

\end{proof}\

4) Assume that the vertex of $h$ has length $q>0$, and is  a
$p$-tuple $\{A_0,A_1,\cdots, A_{p-1}\}$ with $l$ nonzero  elements.
We  prove that the  vertex can be changed to a nonzero $p$-tuple
$\{\tilde{A_0},\tilde{A_1},\cdots, \tilde{A_{p-1}}\}$ with the least
$m$ elements change, such that $\tilde{A_0}=\tilde{A_1}=\cdots=
\tilde{A_{p-1}}$.

\begin{proof}\

Assume that
 $$A_i=\left(\begin{array}{c}a_{1i}\\ a_{2i}\\ \vdots\\
a_{p^qi}\end{array}\right), 0\le i<p.\ \ A=\left(\begin{array}{c}a_{10}\ \ a_{11}\ \ \cdots\ \ a_{1,p-1}\\
a_{20}\ \ a_{21}\ \ \cdots\ \ a_{2,p-1}\\
 \ \ \vdots\\
a_{p^q,0}\ a_{p^q,1}\ \cdots\ a_{p^q,p-1}\end{array}\right).$$

Suppose that the number of nonzero elements in the $j_0$th row of
$A$ is the maximum. Then change all  zero elements in the $j_0$th
row to nonzero elements.

For the $j$th ($j\ne j_0$) row of $A$, if the number of nonzero
elements is greater than the number of zero elements, then change
all  zero elements in the $j$th row to nonzero elements. Otherwise,
change all  nonzero elements in the $j$th row to zero elements.

Let $m$ be the number of all elements changed from
$\{A_0,A_1,\cdots, A_{p-1}\}$ to $\{\tilde{A_0},\tilde{A_1},\cdots,
\tilde{A_{p-1}}\}$. Then one can easily prove that $m$ is the
smallest number of changes such that
$\tilde{A_0}=\tilde{A_1}=\cdots= \tilde{A}_{p-1}$.

\end{proof}\

\section{How I became inspired}


\begin{thebibliography}{3}




\bibitem {Ding}
 Ding, C.S.,  Xiao, G.Z.  and Shan, W.J., The Stability Theory of Stream
Ciphers[M]. Lecture Notes in Computer Science, Vol.561. Berlin/
Heidelberg, Germany: Springer-Verlag, 1991,85-88.

\bibitem {Etzion}
Etzion T., Kalouptsidis N., Kolokotronis N., Limniotis K. and
Paterson K. G., Properties of the Error Linear Complexity Spectrum,
IEEE Transactions on Information Theory, 2009, 55(10): 4681-4686.

\bibitem {Games}
Games, R.A., and Chan, A.H., A fast algorithm for determining the
complexity of a binary sequence with period $2^n$. IEEE Trans on
Information Theory, 1983, {29} (1):144-146.

\bibitem {Fu}
Fu F.,  Niederreiter H., and  Su M., The characterization of $2^n$-periodic binary sequences with fixed
 1-error linear complexity, In: Gong G., Helleseth T., Song H.-Y., Yang K. (eds.) SETA 2006, LNCS, vol. 4086, 88-103. Springer (2006).




\bibitem {Han}
Han Y. K., Chung J. H., and Yang K., On the $k$-error linear
complexity of $p^m$-periodic binary sequences. IEEE Transactions on
Information Theory, 2007, 53(6): 2297-2304.

%


%

\bibitem {Kurosawa}
Kurosawa K., Sato F., Sakata T. and Kishimoto W., A relationship
between linear complexity and $k$-error linear complexity. IEEE
Transactions on Information Theory, 2000, 46(2): 694-698.


\bibitem {Lauder}
Lauder A. and Paterson K., Computing the error linear complexity
spectrum of a binary sequence of period $2^n$. IEEE Transactions on
Information Theory, 2003, 49(1):273-280.



\bibitem {Meidl2002}
 Meidl W. and Niederreiter H., Linear complexity k-error linear complexity,
and the discrete Fourier transform, J. Complexity, 2002,  18:87-103.

\bibitem {Meidl}
Meidl W., How many bits have to be changed to decrease the linear
complexity?, Des. Codes Cryptogr., 2004, 33:109-122.


\bibitem {Meidl2005}
Meidl W., On the stablity of $2^n$-periodic binary sequences. IEEE
Transactions on Information Theory, 2005, 51(3): 1151-1155.

\bibitem {Rueppel}
Rueppel R A. Analysis and Design of Stream Ciphers. Berlin:
Springer-Verlag, 1986, chapter 4.


\bibitem {Stamp}
Stamp, M., and  Martin, C. F., An algorithm for the $k$-error linear
complexity of binary sequences with period $2^{n}$, IEEE Trans.
Inform. Theory, 1993, {39}:1398-1401.


\bibitem {Wei}
Wei, S. M., Xiao, G. Z., and Chen, Z., A fast algorithm for determining
the minimal polynomial of a sequence with period $2p^n$ over
$GF(q)$, IEEE Trans on Information Theory, 2002,
{48}(10):2754-2758.


\bibitem {Xiao}
Xiao, G. Z., Wei, S. M., Lam K. Y., and Imamura K., A fast algorithm for
determining the linear complexity of a sequence with period $p^n$
over $GF(q)$. IEEE Trans on Information Theory, 2000,{ 46}:
2203-2206.


\bibitem {Zhou}
 Zhou, J. Q.,  On the $k$-error linear complexity of sequences with period 2$p^n$ over GF(q),
 Des. Codes Cryptogr., 2011, 58(3)279-296.

\bibitem {Zhou12}
Zhou, J. Q., A counterexample concerning the 3-error linear
complexity of $2^n$-periodic binary sequences, Des. Codes Cryptogr.,
2012,64(3):285-286.

\bibitem {Zhou_Liu}
 Zhou, J. Q.,  Liu, W. Q.,
 The $k$-error linear complexity distribution for $2^n$-periodic binary
 sequences,  Des. Codes Cryptogr., 2013, http://link.springer.com/article/10.1007/s10623-013-9805-8.



\bibitem {Zhou_Liu2013}
 Zhou, J. Q.,  Liu, W. Q.,
 On the  $k$-error linear complexity for $2^n$-periodic binary sequences via Cube Theory, 2013, http://arxiv.org/abs/1309.1829

\bibitem {Zhu}
Zhu, F. X. and Qi, W. F., The 2-error linear complexity of
$2^n$-periodic binary sequences with linear complexity $2^n$-1.
Journal of Electronics (China), 2007,24(3): 390-395.


\end{thebibliography}
\end{document}